\documentclass[letterpaper]{article} % DO NOT CHANGE THIS
\usepackage{aaai2026}
\nocopyright% DO NOT CHANGE THIS
\usepackage{times}  % DO NOT CHANGE THIS
\usepackage{helvet}  % DO NOT CHANGE THIS
\usepackage{courier}  % DO NOT CHANGE THIS
\usepackage[hyphens]{url}  % DO NOT CHANGE THIS
\usepackage{graphicx} % DO NOT CHANGE THIS
\usepackage{natbib}  % DO NOT CHANGE THIS AND DO NOT ADD ANY OPTIONS TO IT
\usepackage{caption} % DO NOT CHANGE THIS AND DO NOT ADD ANY OPTIONS TO IT
\usepackage{algorithm}
\usepackage{algorithmic}
\usepackage{booktabs}
\usepackage{adjustbox}
\usepackage{array}
\usepackage{newfloat}
\usepackage{listings}
\DeclareCaptionStyle{ruled}{labelfont=normalfont,labelsep=colon,strut=off} % DO NOT CHANGE THIS
\floatstyle{ruled}
\newfloat{listing}{tb}{lst}{}
\floatname{listing}{Listing}
\usepackage{color} % (if used in text)
\usepackage[pdftex,hidelinks]{hyperref}
\usepackage{xcolor}
\definecolor{GoogleBlue}{HTML}{4285F4}
\definecolor{GoogleRed}{HTML}{EA4335}
\definecolor{GoogleYellow}{HTML}{FBBC05}
\definecolor{GoogleGreen}{HTML}{34A853}
\newcommand{\DATASET}{%
  \texttt{\textsf{{\color{GoogleBlue}G}%
  {\color{GoogleRed}o}%
  {\color{GoogleYellow}o}%
  {\color{GoogleBlue}g}%
  {\color{GoogleGreen}l}%
  {\color{GoogleRed}e}}TrendArchive}%
}
\usepackage{svg}

\title{\DATASET{}:\\A Year-Long Archive of Real-Time Web Search Trends Worldwide}

\author {
    Aleksandra Urman\textsuperscript{\rm 1},
    Anikó Hannák\textsuperscript{\rm 1},
    Joachim Baumann\textsuperscript{\rm 2}
}
\affiliations {
    \textit{Accepted at ICWSM 2026}\\
    \vspace{0.3cm}
    \textsuperscript{\rm 1}Social Computing Group, Department of Informatics, University of Zurich, Switzerland\\
    \textsuperscript{\rm 2}Stanford NLP Group, Computer Science Department, Stanford University, USA\\
    urman@ifi.uzh.ch, hannak@ifi.uzh.ch, joachimbaumann@stanford.edu
    \vspace{0.3cm}
    \\
    \begin{tabular}{c} % A single centered column
    \raisebox{-0.2\height}{\includegraphics[height=1em]{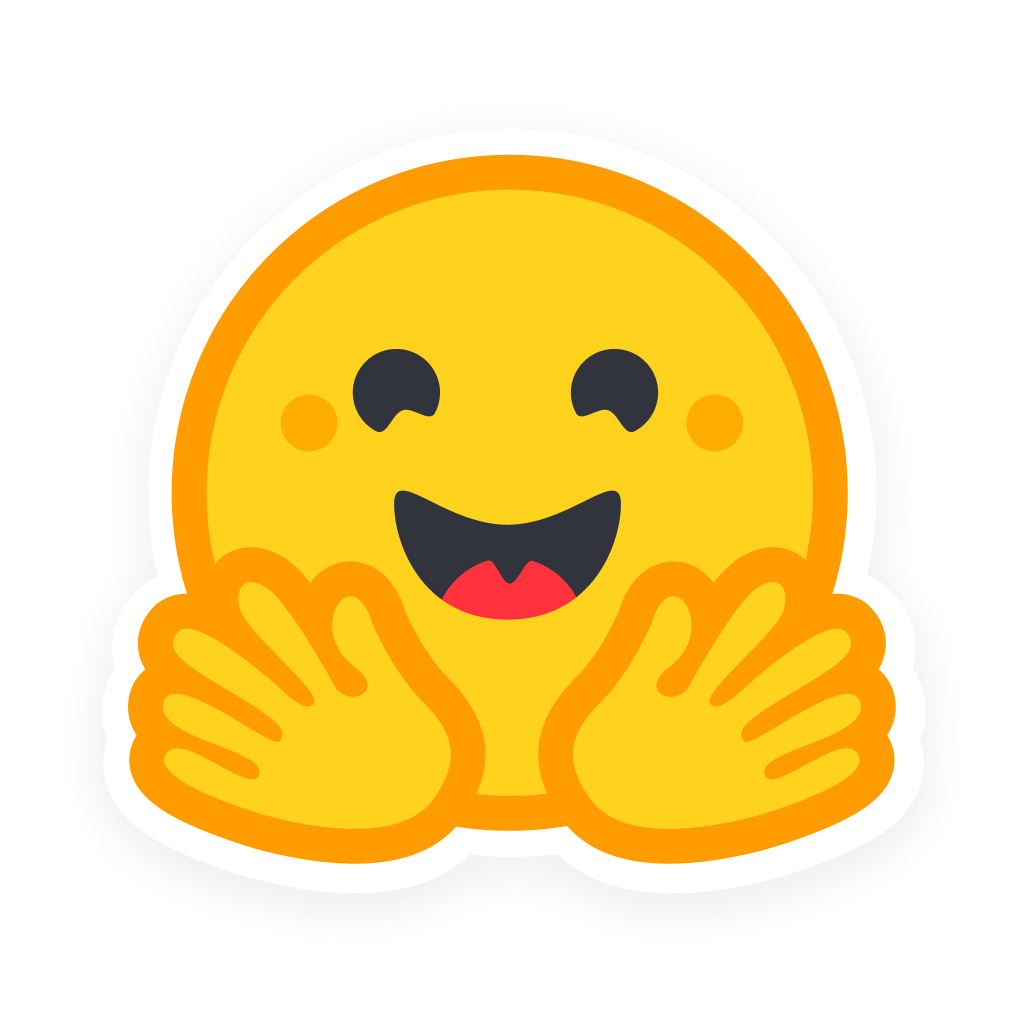}} \href{https://huggingface.co/datasets/aurman/GoogleTrendArchive}{Data} \hspace{0.3cm}
    \raisebox{-0.2\height}{\includegraphics[height=1em]{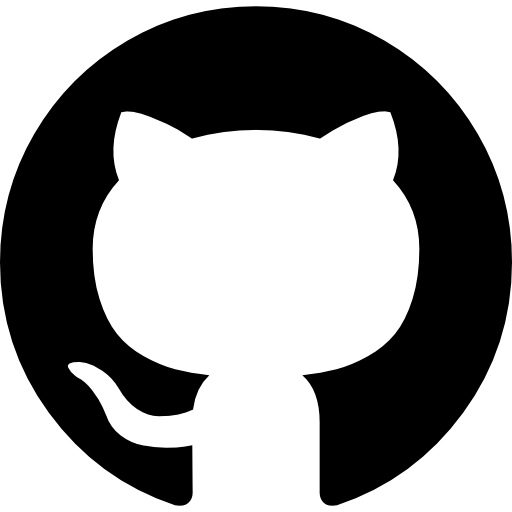}} \href{https://github.com/aurman21/googletrendarchive}{Code}
    \end{tabular}
}

\usepackage{bibentry}
\begin{document}

\maketitle

\begin{abstract}
\DATASET{} is a comprehensive archive of Google Trending Now data spanning over one year (from November 28, 2024 to January 3, 2026) across 125 countries and 1,358 locations. Unlike Google Trends, which requires specifying search terms in advance, Trending Now captures search queries experiencing real-time surges, offering a way to inductively discover trending patterns across regions for studying collective attention dynamics. However, Google does not provide historical access to this data beyond seven days. Our dataset addresses this gap by presenting an archive of Trending Now data. The dataset contains over 7.6 million trend episodes. Each record includes the trend identifier, search volume bucket, precise timestamps, duration, geographic location, and related query clusters. This dataset, among other, enables systematic studies of information diffusion patterns, cross-cultural attention dynamics, crisis responses, and the temporal evolution of collective information-seeking at a global scale. The comprehensive geographic coverage facilitates fine-grained cross-country or cross-regional comparative analyses.
\end{abstract}

\section{Introduction}
\begin{figure}[h!]
\centering
\includegraphics[width=0.8\columnwidth]{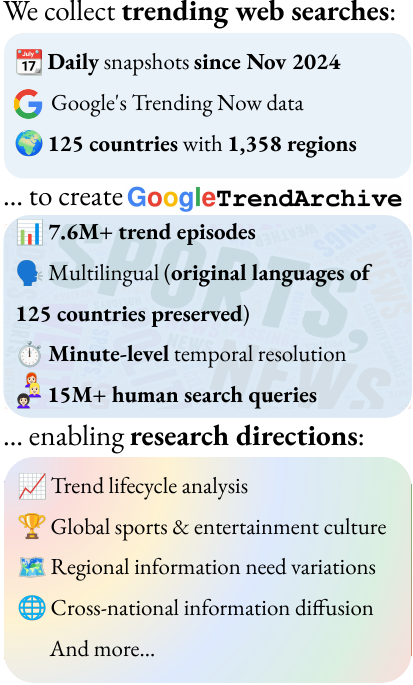}
\caption{\textbf{We release the \DATASET{} dataset}. We collected daily exports of Google's Trending Now data across \textbf{125 countries} and \textbf{1,358 total geographic locations}, creating a \textbf{multilingual} archive of \textbf{7.6M+ trend episodes} spanning \textbf{over one year} (Nov 2024--Jan 2026). Unlike Google Trends, which requires pre-specified queries, \textbf{this dataset enables inductive discovery of what captured collective attention globally}. This supports research on cross-national information diffusion, crisis communication, and cultural variation in real-time search behavior.}
\label{fig:figure1}
\end{figure}

Search engines are among the primary gateways to online information and the key tool users turn to for information-seeking tasks \cite{gleason_google_2023,nielsen_news_2016}. Due to this, web search behavior can be used to evaluate what information internet users care about at a given moment. For this reason, search trends have proven valuable for studying diverse phenomena including: (mental) health-related topics and epidemics \cite{ginsberg_detecting_2009,lazer_parable_2014,nuti_use_2014,zhao_mental_2022,hu2025auditing}; financial and cryptocurrency markets \cite{aslanidis_link_2022,carriere-swallow_nowcasting_2013,preis_quantifying_2013}; economic uncertainty \cite{castelnuovo_google_2017}; unemployment \cite{adu_var_2023}; energy consumption \cite{fu_using_2022}; popularity of specific athletes \cite{malagon-selma_measuring_2023}; fashion trends \cite{skenderi_well_2024}, and a multitude of other topics \cite{jun_ten_2018}. However, existing publicly available resources for studying trending online searches such as Google Trends have a fundamental limitation: researchers must specify in advance what to look for. Specifically, Google Trends allows querying predetermined terms retrospectively, but provides no way to \textit{inductively discover} what actually captured attention across populations without pre-specifying search terms the way Trending Now does.

Addressing this limitation, we present a historical dataset of Google's \textbf{Trending Now} data collected daily over more than one year across all available geographic locations. Trending Now, rolled out to over 100 countries in August 2024 \cite{google_3_2024}, displays search queries experiencing rapid surges in interest---as determined by Google compared to the predicted baseline---in the current moment, refreshing approximately every 10 minutes \cite{google_3_2024}. While anyone can view \textit{current} trends on Google's Trending Now website, the company provides \textit{no historical archive} of such data beyond the last 7 days. Our dataset makes such historical information accessible for researchers, documenting what search queries trended, when the trends started and ended, where they appeared, and how large the surges were. With minimal coverage gaps (approximately 14 days total, concentrated in September 2025, due to technical collection issues), this data enables systematic analysis of collective attention patterns that are otherwise observable only in the moment. Existing archives like the Wayback Machine do not facilitate access to such past data. While Google Trends allows researchers to verify whether specific events they already know about generated search interest, this dataset enables the discovery of what events, topics, and information needs actually emerged as trends across different populations without requiring advance specification. Furthermore, the dataset's comprehensive geographic coverage addresses a critical limitation in computational social science and web search-focused research, where English-language and Western-centric data sources have historically dominated \cite{septiandri_weird_2024}. By capturing trending searches across 125 countries (and a total of 1358 locations, as some countries include data on a regional level), this dataset supports more globally representative studies of collective attention and information diffusion.

This dataset \textbf{facilitates several directions of research}. First, it allows for comparative analyses of information consumption patterns across different geographic contexts, for example examining how attention to global events spreads through various regions or which local phenomena capture attention within specific locations. Second, the high temporal resolution permits temporally fine-grained investigation of the trend lifecycles. Third, researchers can examine the interplay between trending topics and external events across time and locations. Finally, the data also captures the actual search queries people use when seeking information about breaking events, enabling cross-linguistic and cross-geographic comparative analyses of query formulations about the same events.

In the following sections, we first document the dataset structure, the data collection and preprocessing procedures, and then provide a high-level analysis of the data.

%\section{Trending Now vs.\ Google Trends}
%Google Trends and Trending Now serve different purposes and provide different types of data. Google Trends is a retrospective query tool where researchers specify search terms and receive normalized time series data (indexed 0-100) showing relative interest over time \cite{noauthor_faq_nodate}. It enables verification of hypotheses about public interest in known phenomena but requires researchers to know what to look for in advance. Google's 

%Critically, while anyone can visit the Trending Now website to see the current trends, going back up to 7 days, Google does not provide \textbf{public access to historical Trending Now data} beyond that. The information is available only temporarily on the live website, with no archive or API for retrospective access.  Our dataset bridges that gap by systematically collecting daily snapshots over more than one year, creating a retrospective record of what search queries trended, as well as when and where, and for how long they trended. This encompasses historical information that is otherwise not available for research purposes. This dataset thus makes accessible a systematic record of what actually captured collective attention across diverse global Google users over an extended period of time.

\section{\DATASET{} Collection Pipeline}
We developed an automated scraper using Playwright in Python \cite{noauthor_microsoftplaywright-python_2026} to systematically collect Trending Now data from Google's website. The scraper navigates to the Trending Now interface for each location, corresponding to the ``Past 24 hours'' timeframe, and downloads the data via the built-in CSV export function. The collection script and location list are released alongside the dataset to enable verification and possible extension of the data collection.

Data collection began on November 28, 2024, and is ongoing. The scraper initiates daily at 9:00 AM Central European Time and proceeds sequentially through all available locations. Due to processing time variations, the exact collection timestamp for any given location may vary by several minutes from day to day, but all locations are collected within the same daily cycle. The currently released dataset includes all data collected through January 3, 2026, with regular updates planned thereafter, subject to continued infrastructure availability.

We identified 1,358 available geographic locations through manual inspection of the Trending Now interface prior to the data collection. These locations comprise 125 countries plus sub-national regions for select countries---that is, all national and sub-national locations for which Google provides Trending Now data. The complete location list with official codes and names is released along with the dataset (see ``Trends\_LocationList.csv"). As location coverage corresponds to Google's geographic taxonomy, it may change over time as Google adds or removes regions from the Trending Now service. Our data collection has not been affected by such changes so far.

The dataset experienced coverage gaps totaling approximately 14 days due to server outages during the collection period. These gaps occurred when the infrastructure hosting the collection script became unavailable, and outages were not immediately detected. Data validation during collection included verification that the saved CSV files matched the expected format and were non-empty. For sub-national regions with lower search volumes, some daily files are legitimately empty when no trends met Google's detection threshold for that location on that day. We verified this behavior through manual spot-checks comparing empty files against the Trending Now website, confirming that empty files accurately reflected the absence of detected trends rather than collection errors.

It is worth noting that on the July 24, 2025---7 months after our data collection began---Google launched a Google Trends API \cite{google_google_nodate} which facilitates both, access to the historical trends and also the currently trending data. However, the API is available only to a ``very limited number of testers'' \cite{google_introducing_2025}. At the time of writing, we do not have access to this API. For this reason, we have been continuing the data collection using the original Playwright script. While API access would allow others to systematically archive daily trending data, the only such archive known to us \cite{aluoch_daily_nodate} has two main limitations compared to our dataset. First, the data collected by \citet{aluoch_daily_nodate} comprises only the US. Second, this dataset only contains the data starting 19th of September 2025.

\section{\DATASET{} structure and content}

\subsection{Raw Data Collection}

The raw dataset consists of daily snapshots collected across all available locations in Trending Now, spanning over one year. Each snapshot captures trends active at or finished within 24 hours of the time of the collection.
%, with data originally organized as individual CSV files---one file corresponding to one day in one location.
%The collection covers over one year, starting on November 28, 2024, with minimal gaps totaling around 14 days due to technical issues. 

\textbf{Location information} and the \textbf{collection date} are embedded in the file structure, with separate files for each geographic region following the format {\fontsize{7.6}{9.6}\selectfont \verb|trending_{LOCATION_CODE}_{TIMEFRAME}_{TIMESTAMP}.csv|}. Location codes correspond to Google's geographic taxonomy, covering countries and, in some cases, sub-national regions (e.g., US states). The daily files for each location are organized in separate folders, with one folder per location.

Each raw CSV file contains six fields provided by Google's Trending Now system:

The \textbf{(1) trend identifier} represents the search query or cluster of related queries that experienced a surge in search interest. This may be a single query or a representative term---as determined by Google---for a cluster of related searches that Google's system identified as referring to the same underlying topic. For example, the following comma-separated queries ``\textit{man united vs bodø/glimt, manchester united - bodø/glimt, manchester united, man utd, man united, man u, bodo glimt, uefa, manchester united f.c. vs bodø/glimt lineups, where to watch manchester united f.c. vs bodø/glimt, rasmus højlund, manchester united f.c. vs bodø/glimt}'' correspond to the trend identifier ``\textit{man united vs bodø/glimt}'' and represent different ways in which users in the US on November 28, 2024, searched for information about a UEFA Europa League match on that day.

The \textbf{(2) search volume field} provides an approximate number of actual Google searches. Google describes this as ``bucketed traffic for the trend within the selected timeframe"~\citep{google_trends_help}. These categories appear as ranges such as ``100+", ``200+", ``500+", ``50K+", ``100K+", ``200K+", ``500K+", ``2M+", and so forth. Google groups actual traffic into these predetermined buckets rather than providing exact numbers.

Precise temporal information is captured through \textbf{(3) start and (4) end timestamps, recorded with timezone information}. These timestamps indicate when Google's forecasting engine first flagged the trend as emerging and when it determined the trend had ended, returning to baseline search levels. %This fine-grained temporal data enables the analysis of trend lifecycles, which range from minutes to hours to days depending on the nature of the underlying event or topic.

The \textbf{(5) trend breakdown field} lists the specific search queries that compose the trend cluster. When multiple related queries surge simultaneously, Google's system groups them together, and this field lists all the variants of the related queries as displayed in Trending Now---as shown in the ``\textit{man united vs bodø/glimt}'' trend example above. It is worth noting that we found no publicly available information on how the queries are grouped or otherwise preprocessed by Google---a limitation necessary to take into account when interpreting the query data.

Each record includes an \textbf{(6) explore link} pointing to the corresponding Google Trends page for the primary query, enabling verification and further investigation of individual trends through Google's standard Trends interface, including retrospective analysis of historical data.

\subsection{Processed Dataset Release}

To facilitate analysis, we provide a consolidated, processed version of the dataset as a single CSV file. This version addresses several challenges inherent in the raw daily snapshots: (1) the same trending topic may appear in multiple consecutive daily files if it remained active across days, requiring deduplication; (2) timestamp parsing and standardization across timezones; (3) calculation of trend durations; and (4) correction of data quality issues. The complete processing pipeline is described below and the preprocessing code is released along with the dataset.

\subsubsection{Data Loading and Standardization}

We loaded all CSV files across all geographic locations, extracting the collection date from each filename (format: YYYYMMDD) and standardizing column names to lowercase with underscores. Each record was tagged with its source location and collection date to preserve provenance. The combined raw data contained 7,691,790 trend records.

\subsubsection{Field Parsing}

We parsed three key data types from the raw files:
\begin{itemize}
    \item \textbf{Search volume buckets}: We extracted the numeric lower bound of each categorical range (e.g., ``500+'' → 500, ``50K+'' → 50,000, ``2M+'' → 2,000,000) to enable quantitative analysis, while retaining the original categorical values.

    \item \textbf{Timestamps}: We removed timezone suffixes (e.g., ``UTC+1", ``UTC-5") and parsed all timestamps to UTC using the format ``Month day, year at hour:minute:second AM/PM'' (e.g., ``January 1, 2025 at 2:00:00 AM"). This standardization ensures temporal consistency across all locations.

    \item \textbf{Query counts}: We counted the number of individual search queries in each trend cluster by parsing the comma-separated values in the trend breakdown field.
\end{itemize}

\subsubsection{Episode-Based Deduplication}

Because we collected daily snapshots, the same trending topic could appear in multiple consecutive files if it remained active across days. Simply concatenating all files would artificially inflate trend counts. To address this, we implemented episode-based deduplication:

\textit{Episode identification:} For each unique combination of trend identifier and location, we sorted all occurrences chronologically by start time. We defined a new ``episode'' when the start time differed by more than 1 hour from the previous occurrence's start time. This approach correctly handles cases where the same trend appears multiple times in daily snapshots (e.g., once with a missing end time, once with a complete timestamp), ensuring they are merged into a single episode. The 1-hour threshold distinguishes genuinely distinct trending episodes from duplicate observations of the same continuous trend.

\textit{Episode aggregation:} For each episode, we collapsed multiple daily observations into a single record by selecting:
\begin{itemize}
\item Earliest start time across all occurrences (when the trend first emerged)
\item Latest end time across all occurrences (when the trend finally ended)
\item Maximum search volume bucket observed during the episode
\item First observed trend breakdown and explore link
\item Metadata: number of distinct days observed, total raw occurrences, date range
\end{itemize}

This approach preserves the true temporal extent of each trending episode while eliminating duplicate counting. For example, if a major news event trended continuously from Monday through Wednesday, appearing in three daily snapshot files, the processed dataset contains a single episode record spanning those three days.

\subsubsection{Duration Calculation and Quality Corrections}

We calculated trend duration as the time difference between start and end timestamps. However, inspection of the data revealed two quality issues requiring correction:

\textit{Reversed timestamps:} In approximately 0.001\% of records (76 cases), the end timestamp preceded the start timestamp, likely due to timezone handling inconsistencies in Google's system. We corrected these by swapping the start and end timestamps, flagging such records with the boolean field \texttt{times\_were\_swapped}.

\textit{Missing end timestamps:} Approximately 14.8\% of trend episodes lacked end timestamps, typically representing ongoing trends at the time of collection or cases where Google's system did not record a clear end point. For these cases, we estimated the end time as 23:59:59 UTC on that day (assuming the trend ended by day's end, as otherwise it would be recorded in the next daily snapshot).

All estimated durations are flagged with the boolean field \texttt{duration\_is\_estimate} to enable sensitivity analyses. Researchers can choose to exclude estimated durations, use them as upper bounds of trend duration, or employ alternative estimation strategies.

\subsection{Dataset Summary Statistics}

The \textbf{final processed dataset contains 7,639,695 unique trend episodes across 1,339 geographic locations, spanning November 28, 2024 to January 3, 2026}. Table~\ref{tab:dataset_summary} presents key statistics characterizing the dataset.

\begin{table}[h]
\centering
\begin{adjustbox}{max width=\columnwidth}
\begin{tabular}{lr}
\toprule
\textbf{Metric} & \textbf{Value} \\
\midrule
\multicolumn{2}{l}{\textit{Coverage}} \\
Total trend episodes & 7,639,695 \\
Unique locations\textsuperscript{*} & 1,339 \\
Date range & Nov 28, 2024 -- Jan 3, 2026 \\
Total N of queries & 15,394,099 \\
N of unique queries & 1,654,056 \\
\midrule
\multicolumn{2}{l}{\textit{Duration statistics (hours)}} \\
Minimum & 0.2 \\
Maximum & 47.8 \\
Mean & 7.2 \\
Median & 3.3 \\
\midrule
\multicolumn{2}{l}{\textit{Data quality}} \\
Episodes with estimated durations & 1,127,320 (14.8\%) \\
Episodes with corrected timestamps & 76 \\
\bottomrule
\end{tabular}
\end{adjustbox}
\vspace{0.5em}
{\tiny \textsuperscript{*}19 small regional locations had trend files successfully scraped but all of those were empty\\\vspace{-0.3cm}due to the low search volume due to the locations' small size. Those are excluded here.}
\caption{\textbf{Summary statistics} for the processed dataset.}
\label{tab:dataset_summary}
\end{table}

The median trend duration of 3.3 hours reflects the typical lifecycle of trending topics, while the maximum duration of 47.8 hours indicates that some topics (such as major ongoing news events) can sustain elevated search interest over multiple days. The 14.8\% of episodes with estimated durations represents trends that were ongoing at the time of collection or lacked explicit end timestamps from Google's system.

\subsection{Example Record}

Table~\ref{tab:example_record} shows an example record from the processed dataset. This trend represents search activity around a UEFA Europa League match, captured in the United States on November 28, 2024. The trend breakdown has 12 distinct query variants (from team abbreviations like ``man utd'' to specific searches like ``where to watch manchester united f.c. vs bodø/glimt"), and the 11.5-hour duration indicates sustained public interest from pre-match through post-match discussion. The \texttt{total\_occurrences} value of 2 indicates this episode was observed in two daily snapshot files but merged into a single episode as both snapshots corresponded to the same trend, based on the query pattern, location, and trend start time matching. \textbf{The Final Dataset Schema detailing the descriptions of each field in the dataset is available in the Appendix \ref{appendix:schema}.}

\begin{table}[h]
\centering
\small
\begin{adjustbox}{max width=\columnwidth}
\begin{tabular}{ll}
\toprule
\textbf{Field} & \textbf{Value} \\
\midrule
trends & man united vs bodø/glimt \\
search\_volume & 100K+ \\
search\_volume\_lower & 100000 \\
start\_time & 2024-11-28 20:00:00 UTC \\
end\_time & 2024-11-29 07:30:00 UTC \\
duration\_hours & 11.5 \\
n\_queries & 12 \\
trend\_breakdown & man united vs bodø/glimt, manchester\\
& united - bodø/glimt, manchester united,\\
& man utd, man united, man u, bodo glimt, \\
& uefa, [...] \\
location & US \\
collection\_date & 2024-11-29 \\
n\_days\_observed & 1 \\
total\_occurrences & 2 \\
duration\_is\_estimate & FALSE \\
times\_were\_swapped & FALSE \\
\bottomrule
\end{tabular}
\end{adjustbox}
\caption{\textbf{Example trend record} from the processed dataset.}
\label{tab:example_record}
\end{table}

\subsection{Data Availability}

Both the raw daily snapshot files and the processed consolidated dataset are available for download at \url{https://doi.org/10.57967/hf/7531}. The raw files preserve the original structure and content as captured from Google's Trending Now system, while the processed dataset provides a ready-to-analyze format. We also provide the complete \href{https://github.com/aurman21/googletrendarchive}{data processing code} (R) to ensure full reproducibility and to enable researchers to implement alternative processing strategies if desired. In addition, we provide a Datasheet \cite{gebru2021datasheets} accompanying the dataset, available at the same DOI.

%\section{Coverage Gaps}

\section{High-level Analysis of the Data}
In this section, we present high-level analyses of the data. All analyses are based on the processed---as described above---dataset of 7,639,695 trend episodes, unless specified otherwise.

\subsection{Descriptive Statistics: Number of Trends, Queries per Trend, Trend Duration, and Average Search Volume per Trend}
Table \ref{tab:countrystats} presents country-level descriptive statistics for the top 30 countries by total trend count. The United States showed the highest trending activity with 175,143 total trends and 67,779 unique trends, followed by Japan (146,587 total trends) and the United Kingdom (116,113 total trends). The average number of queries per trend varies from 1.70 (South Korea) to 3.2 (Egypt), with most countries clustering between 2-3 queries per trend. Search volume averages are highest in India (11,774), the United States (10,957), and Brazil (9,239). The cross-national variation observed across variables may reflect differences in population size, internet penetration, search behavior patterns, topic diversity, or Google's trend detection thresholds across markets, though we cannot definitively attribute these patterns to any single factor without additional data on Google's algorithmic parameters by region.

The systematic difference between actual (3-6 hours) and estimated (20-26 hours) trend durations is an artifact of our estimation methodology. For trends lacking observed end times, we conservatively assigned 23:59:59 on the last observation day as the end time. This mechanically produces longer durations: a trend first observed at any time on day N and last seen (still ongoing) on day N receives an estimated duration spanning from its start time to end of day N. The tight clustering of estimated durations around 20-26 hours across all countries---with minimal cross-national variation compared to actual durations---confirms this is methodological rather than substantive. These estimated values represent upper bounds (maximum possible durations): trends ended sometime between their last observation and end of that day, but we cannot determine precisely when. \textbf{Researchers should interpret estimated durations as ``lasted at most X hours"} and consider sensitivity analyses excluding estimated durations for duration-dependent analyses.
\begin{table*}[hbt!]
\caption{\label{tab:countrystats}\textbf{Country-level descriptive statistics} (top 30 countries by the N of total trends)}
\centering
\adjustbox{max width=\textwidth}{%
\begin{tabular}[t]{rlrrrrrrrr}
\toprule
Rank & Country & Total Trends & Unique Trends & Total Queries & Avg Queries/Trend & Avg Dur Actual (hrs) & Avg Dur Est (hrs) & Avg Volume\\
\midrule
1 & United States & 175,143 & 67,779 & 468,912 & 2.7 & 3.4 & 22.5 & 10,957\\
2 & Japan & 146,587 & 47,217 & 264,493 & 1.8 & 3.5 & 22.0 & 2,963\\
3 & United Kingdom & 116,113 & 48,205 & 261,915 & 2.3 & 3.6 & 21.6 & 3,840\\
4 & Germany & 99,788 & 44,584 & 198,577 & 2.0 & 3.6 & 24.1 & 4,028\\
5 & Brazil & 94,391 & 39,329 & 285,936 & 3.0 & 3.5 & 25.2 & 9,239\\
\addlinespace
6 & Canada & 90,088 & 41,013 & 193,403 & 2.1 & 3.9 & 23.5 & 2,522\\
7 & France & 87,901 & 37,747 & 183,739 & 2.1 & 3.3 & 23.9 & 3,785\\
8 & India & 85,897 & 36,437 & 223,563 & 2.6 & 4.0 & 24.2 & 11,774\\
9 & Italy & 85,198 & 37,532 & 200,316 & 2.4 & 3.4 & 24.2 & 4,429\\
10 & Spain & 66,189 & 30,498 & 149,606 & 2.3 & 3.5 & 23.1 & 4,465\\
\addlinespace
11 & Türkiye & 65,513 & 29,352 & 179,402 & 2.7 & 4.0 & 22.6 & 6,851\\
12 & Indonesia & 59,298 & 26,904 & 137,912 & 2.3 & 4.1 & 25.1 & 4,619\\
13 & Mexico & 54,931 & 26,968 & 134,024 & 2.4 & 3.8 & 23.4 & 6,609\\
14 & Russia & 50,314 & 24,113 & 98,423 & 2.0 & 5.7 & 23.8 & 2,064\\
15 & Australia & 47,956 & 25,045 & 98,987 & 2.1 & 3.8 & 22.9 & 2,325\\
\addlinespace
16 & Argentina & 44,175 & 20,895 & 106,996 & 2.4 & 3.6 & 24.8 & 5,729\\
17 & Netherlands & 41,598 & 21,575 & 79,095 & 1.9 & 4.6 & 23.2 & 2,153\\
18 & Poland & 39,776 & 19,408 & 81,615 & 2.1 & 3.7 & 23.1 & 3,311\\
19 & Taiwan & 38,294 & 17,486 & 72,783 & 1.9 & 4.9 & 22.2 & 1,879\\
20 & Iran & 33,613 & 14,663 & 74,626 & 2.2 & 4.0 & 23.9 & 2,664\\
\addlinespace
21 & South Korea & 33,054 & 13,961 & 55,735 & 1.7 & 4.9 & 23.4 & 2,507\\
22 & Colombia & 31,382 & 16,139 & 93,121 & 3.0 & 4.0 & 24.6 & 6,983\\
23 & Saudi Arabia & 31,082 & 16,163 & 71,167 & 2.3 & 5.8 & 24.7 & 3,205\\
24 & South Africa & 30,348 & 15,834 & 70,025 & 2.3 & 4.8 & 25.7 & 3,101\\
25 & Chile & 29,700 & 15,006 & 70,594 & 2.4 & 3.9 & 24.4 & 3,754\\
\addlinespace
26 & Egypt & 29,408 & 15,755 & 93,265 & 3.2 & 5.8 & 27.0 & 4,349\\
27 & Peru & 28,171 & 14,482 & 77,526 & 2.8 & 4.4 & 24.7 & 5,618\\
28 & Bangladesh & 27,056 & 14,458 & 53,296 & 2.0 & 5.5 & 26.0 & 3,958\\
29 & Vietnam & 26,503 & 13,169 & 76,078 & 2.9 & 4.3 & 20.7 & 5,116\\
30 & Philippines & 25,950 & 12,767 & 53,952 & 2.1 & 3.9 & 24.0 & 2,351\\
\bottomrule
\end{tabular}%
}
\end{table*}

\subsection{Topical Distribution of Trends}

\paragraph{Top Trends in 2025.}

\begin{figure}[hbt!]
\centering
\includegraphics[width=\columnwidth]{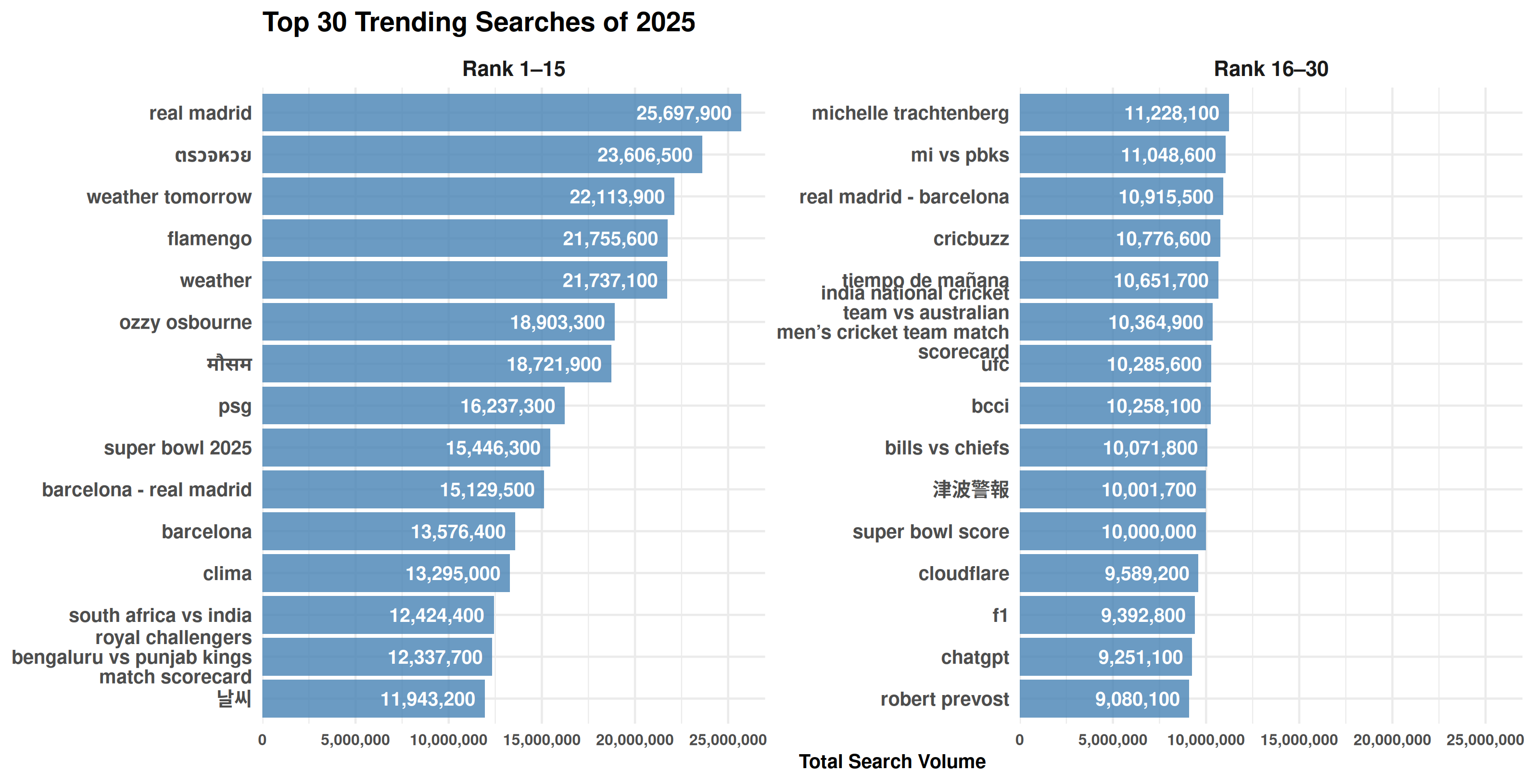}
\caption{\textbf{Top 30 Trends} of 2025 (Country-level data only) by Total Search Volume.}
\label{fig:top30_2025}
\end{figure}
In Figure \ref{fig:top30_2025} we present an overview of the top 30 trends of 2025 (based on country-level data only) by total search volume. \textbf{Most top trends are related to sports events or weather}. This pattern is consistent for most regions.
% we observe in the data more generally, as we describe based on the analyses reported below.

\paragraph{Top Trending Topics by Region.}

\begin{figure}[hbt!]
\centering
\includegraphics[width=\columnwidth]{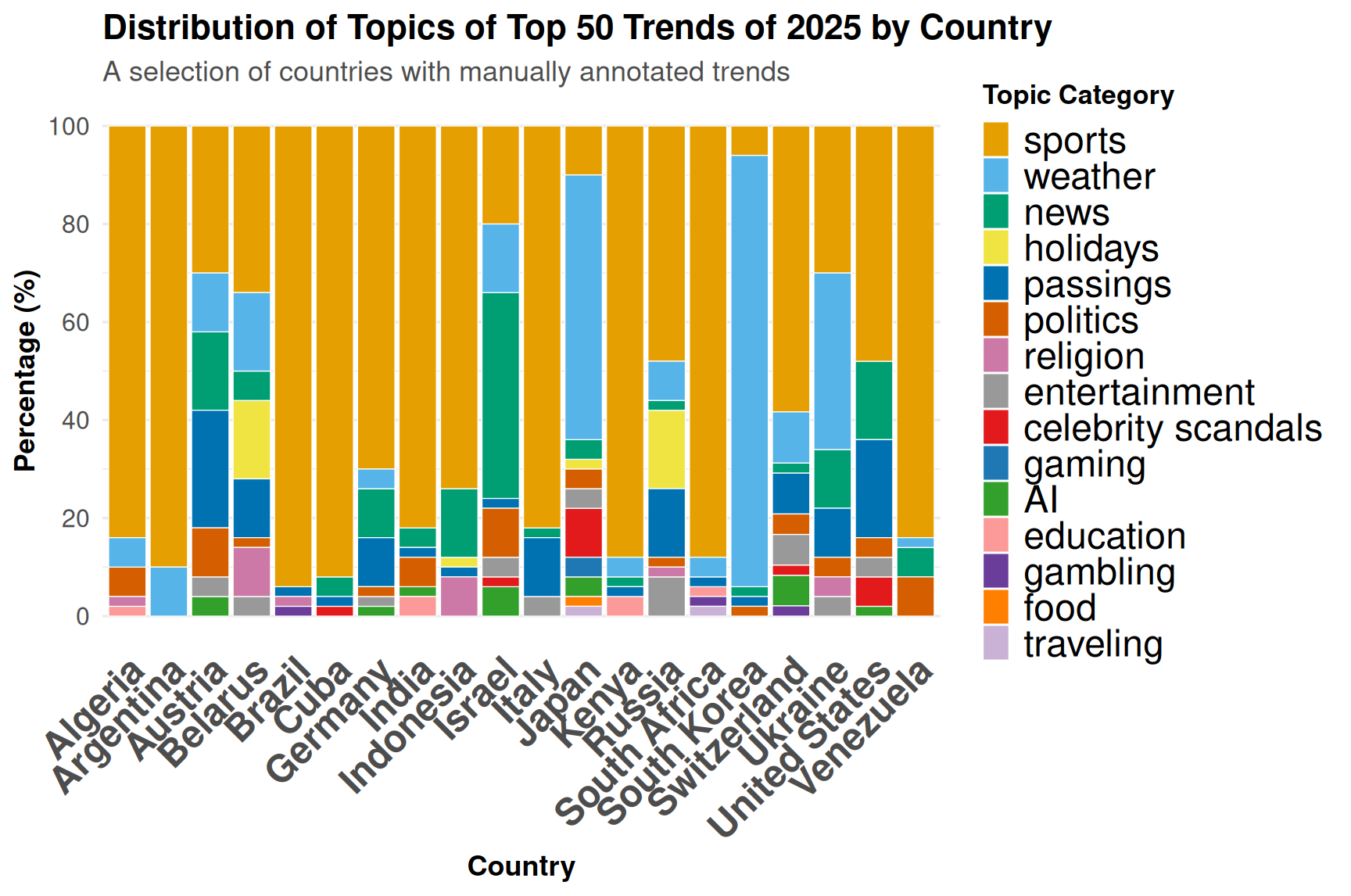}
\caption{\textbf{Distribution of topics} among the top 50 trending searches in 2025 for 20 randomly selected countries. To avoid risks of large-scale automated annotations~\citep{baumann2025llmhacking}, topics were manually annotated by the authors based on trend titles and associated queries. For languages not spoken by the authors, trend titles were automatically translated to English, with web search used to clarify ambiguous terms or culturally specific references.}
\label{fig:topic_distribution}
\end{figure}

\textit{Topic Annotation:}
To characterize the substantive content of trending searches, we manually annotated the top 50 trends from a subset of 20 countries, representing diverse geographic regions and political systems. The authors classified each trend into 15 mutually exclusive topic categories: sports, weather, news, passings (deaths of notable figures), holidays, celebrity scandals, politics, religion, entertainment, gaming, AI, education, food, gambling, and traveling.

For languages not spoken by any of the authors (i.e., Japanese, Arabic, Hebrew, Indonesian, Portuguese), trend titles and associated queries were automatically translated to English using machine translation. When translations were ambiguous or culturally specific, we employed web search to clarify the nature and context of trending topics. This hybrid approach enabled us to conduct reliable classification across linguistically, geographically, culturally and politically diverse contexts.

\textit{Results:}
Figure~\ref{fig:topic_distribution} reveals that sports-related searches dominate trending topics across most countries, often comprising 50-80\% of top trends in most of the selected countries except Austria, Belarus, Israel, Japan, South Korea, and Ukraine. These predominantly relate to ongoing sporting events, with specific sports varying by region: cricket features prominently in India, American football (NFL) dominates in the United States, while football (soccer) is the most common sport searched for in other contexts. Given this dominance, researchers using this dataset for purposes unrelated to sports may wish to filter out sports-related trends to focus on the remaining content categories, which still provide substantial volume and diversity for analysis.

Some exceptions to sports dominance in top trends occur in countries experiencing crises, armed conflicts or natural disasters. In Japan, weather-related trends (largely tsunami warnings) constituted a substantial portion of searches. Notably, in South Korea where the majority of trends also corresponded to weather, they referred to ``mundane'' weather searches, not related to natural disasters. Israel showed elevated news-related searches, predominantly concerning the ongoing armed conflict and the lives and release of hostages. Ukraine also showed heightened attention to news topics, particularly air raid alerts and electricity shutdown notifications. However, otherwise, political topics remain relatively modest in most countries' trending searches, suggesting that while politically significant events may trend, they do not dominate the overall landscape of real-time search behavior. Entertainment, celebrity scandals, and gaming show more variable patterns across countries, possibly reflecting the cultural differences in media consumption patterns.

\subsection{Multi-Country Trends and Global Information Diffusion Patterns}
Table \ref{tab:top30globaltrends} shows top 30 global (i.e., having appeared in multiple countries) trends, ordered by the number of countries they appeared in. For this analysis, we utilized only country-level (not regional) data. As seen in Table \ref{tab:top30globaltrends}, among the top 30 trends that appeared in multiple countries, most correspond to various sporting events and topics. Beyond those, trends related to generative AI tools---ChatGPT, Deepseek, Gemini---stand out. ChatGPT was the only trend having appeared in all 125 countries in the dataset. Additionally, several political trends appeared globally such as those related to Donald Trump or Iran, and trends related to celebrations and holidays such as New Year or Earth Day.

\begin{table}[hbt!]
\caption{\label{tab:top30globaltrends}\textbf{Top 30 Multi-Country Trends} by the Total N of Countries They Appeared In.}
\centering
\adjustbox{max width=\columnwidth}{%
\begin{tabular}[t]{rlrrrr}
\toprule
Rank & Trend & N Countries & N Occurrences & Avg Volume & Max Volume\\
\midrule
1 & chatgpt & 125 & 592 & 18,324 & 2,000,000\\
2 & psg & 124 & 2,406 & 6,841 & 2,000,000\\
3 & real madrid & 124 & 3,664 & 7,615 & 1,000,000\\
4 & la liga & 123 & 2,471 & 2,653 & 200,000\\
5 & barcelona & 122 & 2,440 & 5,922 & 500,000\\
\addlinespace
6 & champions league & 121 & 1,232 & 5,514 & 500,000\\
7 & uefa champions league & 121 & 1,212 & 2,341 & 100,000\\
8 & weather & 121 & 4,439 & 7,207 & 1,000,000\\
9 & new year's eve & 120 & 234 & 28,668 & 2,000,000\\
10 & diogo jota & 119 & 230 & 34,315 & 1,000,000\\
\addlinespace
11 & weather tomorrow & 119 & 2,552 & 11,841 & 1,000,000\\
12 & chelsea & 118 & 1,498 & 3,672 & 200,000\\
13 & new year's day & 118 & 224 & 22,729 & 1,000,000\\
14 & premier league & 118 & 1,187 & 4,098 & 500,000\\
15 & deepseek & 117 & 213 & 5,916 & 500,000\\
\addlinespace
16 & gemini & 116 & 494 & 4,644 & 200,000\\
17 & liverpool & 116 & 1,842 & 5,278 & 1,000,000\\
18 & earth day & 115 & 128 & 15,816 & 500,000\\
19 & hulk hogan & 115 & 152 & 34,663 & 1,000,000\\
20 & psg vs inter miami & 115 & 118 & 2,241 & 20,000\\
\addlinespace
21 & real madrid vs barcelona & 115 & 308 & 6,383 & 100,000\\
22 & barcelona vs real madrid & 114 & 264 & 21,563 & 1,000,000\\
23 & copa del rey & 114 & 692 & 2,572 & 200,000\\
24 & arsenal & 113 & 1,507 & 5,827 & 200,000\\
25 & serie a & 113 & 1,147 & 2,344 & 100,000\\
\addlinespace
26 & europa league & 112 & 942 & 4,052 & 200,000\\
27 & donald trump & 111 & 744 & 1,383 & 100,000\\
28 & iran & 111 & 405 & 4,313 & 200,000\\
29 & ufc & 111 & 1,142 & 9,858 & 1,000,000\\
30 & nba & 110 & 2,189 & 2,581 & 1,000,000\\
\bottomrule
\end{tabular}%
}
\end{table}

\begin{figure*}[hbt!]
\centering
\includegraphics[width=1.8\columnwidth]{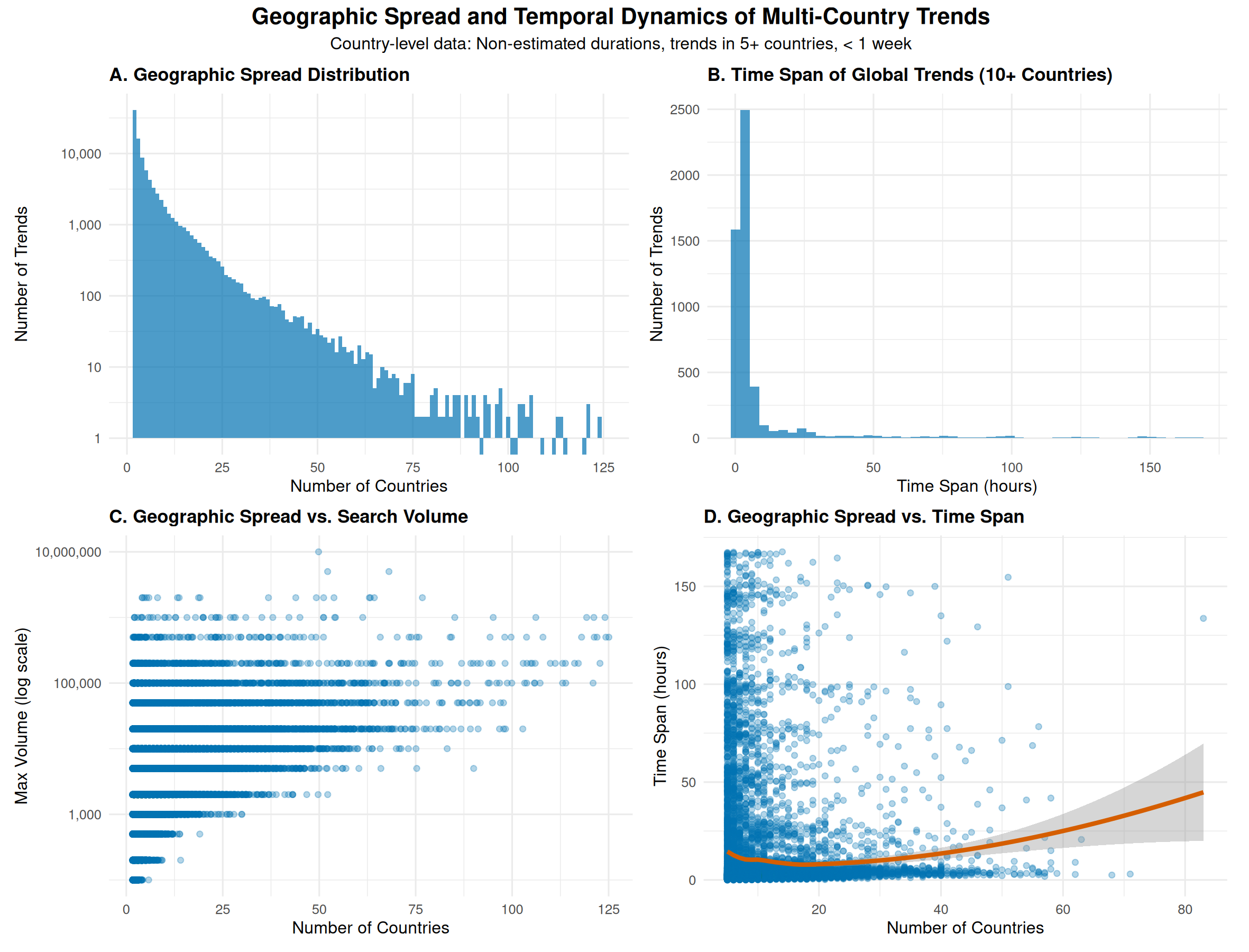}
\caption{\textbf{Geographic spread and temporal dynamics of multi-country trends}; only trend episodes with actual---i.e., not estimated---durations are included.  (A) Distribution of geographic spread showing power-law pattern. (B) Time span distribution for trends in 10+ countries. (C) Positive correlation between geographic spread and search volume (Spearman $\rho$ = 0.43, p $<$ 0.001). (D) Non-monotonic relationship between geographic spread and time span, with medium-spread trends (20-40 countries) showing shorter durations than broadly  global trends (60+ countries). Orange lines: LOESS smoothing; gray shading: 95\% CI. Country-level trends with 5+ countries, $<$ 1 week.}
\label{fig:geospread}
\end{figure*}

To characterize the diffusion patterns of trending searches across countries, we analyzed the relationship between geographic spread (number of countries where a trend appears), temporal persistence (time span between first and last appearance), and search volume. For this analysis, we utilized only trends with actual---i.e., non-estimated,---durations to reduce the noise level in the data. Figure \ref{fig:geospread} presents four complementary views of these dynamics. The distribution of geographic spread (Panel A) follows a power-law-like pattern where 95.1\% of trends remain localized to fewer than 20 countries, 4.8\% of trends reach 20-99 countries, and only 0.01\% achieve truly global reach (100+ countries). The temporal analysis (Panel B) of trends appearing in 10+ countries shows that most cross-border diffusion occurs within 72 hours, suggesting rapid information cascades across national boundaries. Specifically, 91\% of trends that transcend country boundaries achieve that in less than 24 hours; 5.1\% take 24-72 hours; 3.8\% take longer than 72 hours. Furthermore, we observe a positive correlation (Spearman $\rho$ = 0.43, p $<$ 0.001, N = 99,167) between geographic spread and search volume (Panel C): trends reaching more countries tend to generate higher search volumes, though substantial variance exists at all levels of spread. The relationship between geographic spread and time span (Panel D) shows a non-monotonic pattern: trends spreading to 20-40 countries show relatively short durations (often under 20 hours), suggesting rapid but short-lived viral diffusion, while trends achieving broader geographic reach (60+ countries) demonstrate more sustained temporal persistence. This pattern suggests that universal relevance rather than mere virality drives lasting global attention. These high-level findings showcase how our dataset might be relevant for research on cross-national or cross-regional (if regional data was to be used) information diffusion, agenda-setting across borders, and the identification of short-lived ``viral'' versus ``sustained'' global search trends.

\section{Research Directions enabled by \DATASET{}}
\label{sec:research_directions}
%This analysis demonstrates the dataset's value for multiple research domains. The prevalence of sports trends provides opportunities for studying global sports culture, information-seeking around live events, and cross-national patterns in sports fandom. The crisis-driven variations in Japan, Israel, and Ukraine illustrate how the dataset captures real-time public attention during emergencies and armed conflicts, enabling research on crisis communication and information needs during crises. The heterogeneity across topic categories---from celebrity culture to political events to weather phenomena---supports comparative research on collective attention, cultural priorities, and the interplay between global and local concerns in digital information-seeking behavior. Researchers can, for example, leverage the dataset's temporal granularity to examine how trending topics evolve hour-by-hour, providing insights into information diffusion patterns, the lifecycle of viral content, and the rapid shifts in public attention that characterize contemporary digital media environments.

This dataset enables a wide range of research questions across multiple domains of computational social science, information science, and communication science research. We identify several key areas where we believe search data can be particularly useful, as partially exemplified by prior work on similar topics done using Google Trends data.

\subsection{Information Diffusion and Attention Patterns}
The temporal granularity and geographic coverage of this dataset make it particularly suited for studying how information spreads across contexts. Researchers can track how attention to breaking news events spreads across countries/regions, examining the speed and pathways of information diffusion. The dataset enables investigation of which events attract attention globally (capturing attention simultaneously across locations) versus those that remain geographically localized. By analyzing trend lifecycles---from emergence through peak to decay---scholars can identify temporal diffusion and attention patterns associated with different event types: natural disasters, political developments, entertainment news, or sporting events might exhibit distinct attention patterns that can be characterized and modeled \cite{nghiem_analysis_2016,timoneda_spikes_2022,mellon_internet_2014}.

\subsection{Crisis Communication and Public Response Patterns}
The real-time and temporally fine-grained nature of Trending Now data makes it valuable for studying how publics respond to crises and emergencies. By using the queries associated with the crises-related search trends, researchers can analyze the immediate information needs that emerge during natural disasters, public health events, or political crises. The dataset allows for comparison of crisis response patterns across contexts, identification of information gaps that drive search behavior, and tracking of how attention shifts as crises evolve \cite{oswald_i_2026,rovetta_reliability_2021,timoneda_spikes_2022}.

\subsection{Comparative Cultural, Socio-Political, and Media Analyses}
The comprehensive geographic coverage enables systematic comparisons of collective attention patterns across cultural, linguistic, and political contexts. Researchers can examine cross-national or cross-regional attention patterns to specific topics, analyze whether certain issues resonate universally or remain culturally specific, and how political systems shape information-seeking behavior. For instance, scholars might examine how authoritarian and democratic contexts differ in their patterns of political information-seeking, or whether patterns of public attention to certain issues are related to the local media system structures \cite{dancy_global_2024,kupfer_russian_2022}. The dataset also permits the analysis of linguistic and cultural boundaries in information flows---i.e., whether and how different topics spread across these divides. Furthermore, by comparing search trends with social media trends, news coverage, and other information sources, researchers can study how different platforms and media types interact within the broader information ecosystem \cite{nghiem_analysis_2016}. Questions such as whether search leads or follows social media trends, how traditional media events translate into search behavior, and whether certain topics dominate particular platforms can be systematically investigated.

\section{Limitations and Ethical Considerations}
While this dataset offers substantial research potential, users should be aware of several limitations and ethical considerations.
\paragraph{Original Data Characteristics.} Search volume is provided by Google in buckets rather than exact counts, limiting precision in quantitative analyses. Google's trend identification algorithm is proprietary, meaning the exact clustering of related queries is not transparent.
% The trends represent relative surges rather than absolute search volumes, which affects interpretation.
\paragraph{Coverage Gaps.} Technical issues resulted in approximately 14 days of missing data, which may have excluded some events. Geographic availability is determined by Google's system and may change over time. Certain locations may have different data quality or coverage patterns due to how Google's proprietary trend identification algorithms work.
\paragraph{Representativeness.} Search behavior on Google does not represent entire populations uniformly. Digital divides, device access, and search engine preferences mean that Google Trends reflects particular demographic segments more than others. Cross-country comparisons must account for varying internet penetration rates and search engine market share. For instance, while Google holds over 80\% market share in most countries where it is accessible and corresponding Trending Now data is thus present---which excludes, e.g., China,---there are prominent exceptions to this. For instance, in Russia and South Korea Google is currently less popular than local search engines, Yandex and Naver, respectively \cite{statbase_search_2026}. We suggest the researchers take the market share of Google into account when using the dataset and, where applicable and possible, complement the Google data released by us with similar data from locally relevant search engines such as Yandex Wordstat \cite{yandex_yandex_2026} or Naver SearchTrend \cite{naver__2026}.

\paragraph{Privacy and Ethical Considerations.} While trending searches are aggregated public data, researchers should consider the privacy implications of analyzing collective search behavior. Patterns in health-related searches, for instance, might reveal sensitive information about populations. Studies using this data should follow ethical research practices and consider potential harms from publicizing certain attention patterns.

\paragraph{Lessons from Google Flu Trends.}
Google Flu Trends famously attempted to forecast influenza prevalence based on search query volumes but produced increasingly inaccurate estimates over time~\cite{butler2013google,Olson2013}.
The core lessons were that search data should supplement rather than substitute for traditional measurement, and that the relationship between search behavior and real-world phenomena is unstable due to both platform-side changes and shifts in user behavior~\cite{lazer_parable_2014}.
These concerns are also relevant for our dataset.
Google may change how it detects or clusters trending queries at any time.
Furthermore, there is a risk of feedback loops where users search for topics because they are listed as trending~\cite{pagan2023}, for example through search engine optimization driven by the public visibility of trends.

\section{FAIR Principles Compliance}
Our dataset is compliant with the FAIR principles in the following ways:

\paragraph{Findable \& Accessible.} The dataset is freely available and has a persistent DOI via Hugging Face (\url{https://doi.org/10.57967/hf/7531}).
% , and is indexed with standard citation format.

% \paragraph{Accessible.} The dataset is freely available at \url{https://doi.org/10.57967/hf/7531} 

\paragraph{Interoperable.} The dataset consists of CSV files; we also release an accompanying preprocessing R script. No proprietary formats or tools are necessary to access or work with the dataset.

\paragraph{Reusable.} The dataset is released under CC-BY-4.0 with complete methodology documentation, datasheet, and preprocessing code.

\section{Conclusion}
We present \DATASET{}, a year-long archive of Google Trending Now data across 125 countries and 1,358 total locations. This resource enables systematic study of collective attention and real-time information-seeking at global scale. Unlike aggregated retrospective tools, Trending Now captures surges as they happen, offering unprecedented temporal resolution for studying attention dynamics.

The dataset addresses critical limitations in existing web search resources. The comprehensive geographic coverage enables cross-cultural comparisons, and near-complete temporal coverage supports robust time series analysis. We envision applications spanning information diffusion modeling, crisis communication, comparative cultural studies, and computational social science methodology more broadly.

\section*{Acknowledgments}

This work was supported by Swiss National Science Foundation (SNSF) grant P500-2\_235328 (JB) and SNSF Project Grant 215354 (AU and AH).

The authors acknowledge that they have used Claude to edit/generate parts of the code for Figure 3 and Figure 4 (with subplots). The authors have also used Claude to add comments to the data preprocessing code released alongside the manuscript/dataset. The authors have verified that no errors were introduced into the data processing and analysis in the generated code/comments. The authors confirm that they have not used AI-based tools for other parts of the research processes, including the writing of the manuscript.
\bibliography{references}

\appendix

\section{Appendix: Detailed Dataset Schema}
\label{appendix:schema}

The processed dataset contains the following fields:

\paragraph{Core trend information}
\begin{itemize}
\item \texttt{trends}: Trend identifier (representative query or cluster name)
\item \texttt{trend\_breakdown}: Comma-separated list of all queries in the cluster
\item \texttt{n\_queries}: Number of queries in the trend breakdown
\item \texttt{explore\_link}: URL to Google Trends page for this trend
\end{itemize}

\paragraph{Search volume}
\begin{itemize}
\item \texttt{search\_volume}: Original categorical bucket (e.g., ``50K+")
\item \texttt{search\_volume\_lower}: Numeric lower bound of the bucket
\end{itemize}

\paragraph{Temporal information}
\begin{itemize}
\item \texttt{start\_time}: When Google first detected the trend (UTC)
\item \texttt{end\_time}: When Google determined the trend ended (UTC)
\item \texttt{duration\_minutes}: Calculated duration in minutes
\item \texttt{duration\_hours}: Calculated duration in hours
\item \texttt{collection\_date}: Date when first observed in our collection
\item \texttt{first\_collection\_date}: First date the episode appeared (if multi-day)
\item \texttt{last\_collection\_date}: Last date the episode appeared (if multi-day)
\item \texttt{year}, \texttt{month}, \texttt{weekday}: Extracted date components
\end{itemize}

\paragraph{Geographic information}
\begin{itemize}
\item \texttt{location}: Geographic location code (e.g., ``US", ``DE", ``BR")
\end{itemize}

\paragraph{Episode metadata}
\begin{itemize}
\item \texttt{episode\_id}: Unique identifier within each trend-location combination
\item \texttt{n\_days\_observed}: Number of distinct days the trend appeared
\item \texttt{total\_occurrences}: Number of raw daily records collapsed into this episode
\end{itemize}

\paragraph{Data quality flags}
\begin{itemize}
\item \texttt{duration\_is\_estimate}: Boolean indicating whether duration is estimated (missing end time)
\item \texttt{times\_were\_swapped}: Boolean indicating timestamp correction (reversed start/end)
\end{itemize}

\end{document}